\documentclass[pra,aps,preprint,byrevtex,showpacs]{revtex4}

\usepackage{epsfig}

\newcommand {\bra}[2] {\mbox{}_{#2}\langle #1 |} 
\newcommand {\ket}[2] {| #1 \rangle_{#2}} 
\newcommand {\braket}[4] {\mbox{}_{#3}\langle #1 | #2 \rangle_{#4}} 
\newcommand {\eqn}[1] {Eq.~(\ref{#1})} 
 
\newcommand {\fig}[1] {Fig.~\ref{#1}} 
\newcommand {\figwidth} {90mm} 
 
\newcommand {\Sec}[1] {Sec.~\ref{#1}}

\begin{document}

\title{Conditional two mode squeezed vacuum teleportation}

\author{P. T. Cochrane} \email{cochrane@physics.uq.edu.au}

\author{T. C. Ralph}

\author{G. J. Milburn}

\affiliation{Department of Physics, The University of Queensland, 
St.~Lucia, Queensland 4072, Australia}

\date{\today}

\begin{abstract}
We show, by making conditional measurements on the 
Einstein-Podolsky-Rosen (EPR) squeezed vacuum, that one can improve 
the efficacy of teleportation for both the position difference, 
momentum sum and number difference, phase sum continuous variable 
teleportation protocols.  We investigate the relative abilities of the 
standard and conditional EPR states, and show that by conditioning we 
can improve the fidelity of teleportation of coherent states from 
below to above the $\bar{F} = 2/3$ boundary.
\end{abstract}


\maketitle

\section{Introduction}

Over recent times teleportation has shown itself to be a fundamental 
building block in the business of quantum information 
processing~\cite{Bennett:1993:1,Vaidman:1994:1,Furusawa:1998:1,%
Bouwmeester:1997:1,Braunstein:1998:2,Gottesman:1999:2} coming in both 
discrete and continuous 
formulations~\cite{Braunstein:1998:2,Ralph:1998:1,Bennett:1993:1,%
Cochrane:2000:2}.  In continuous variable teleportation the 
entanglement resource is the two-mode squeezed state, or the 
Einstein-Podolsky-Rosen (EPR) state\footnote{Strictly speaking, the 
two-mode squeezed vacuum is not an EPR state, however, in the limit of 
infinite squeezing it tends to the EPR state.  This limit is 
unphysical since infinite squeezing requires infinite energy, but for 
moderate squeezing the two-mode squeezed vacuum displays EPR 
\emph{correlations} and for this reason is often referred to in the 
literature as an EPR state.  We shall follow this convention here.}.  
The quality of teleportation depends upon how squeezed the EPR state 
can be made.  High levels of squeezing are hard to achieve, so other 
techniques for improving teleportation can be considered.  Opatrn\'y 
{\it et al.}~\cite{Opatrny:2000:1} showed that one can improve 
standard continuous variable teleportation by conditioning off 
detection results from very slightly reflective beam splitters 
inserted into each arm of the entanglement resource.  Making such 
conditional measurements selects a sub-ensemble of more highly 
entangled states which can then be used to teleport more effectively.  
From this point of view it is similar to a distillation protocol.  The 
conditioning procedure also gives information on when one should 
attempt to teleport the input state, thereby improving the efficiency 
of teleportation.  In this paper we look at a number of 
generalisations to the original scheme.  In particular we consider the 
relative merits of the conditioned and unconditioned EPR states.  We 
also give an extension to the number difference, phase sum 
teleportation protocol of Milburn and 
Braunstein~\cite{Milburn:1999:1}, and analyze the effectiveness of 
conditioning for improving coherent state teleportation for a range of 
conditions.

\section{Improvement of the entanglement resource}

\begin{figure}
\centerline{\epsfig{file=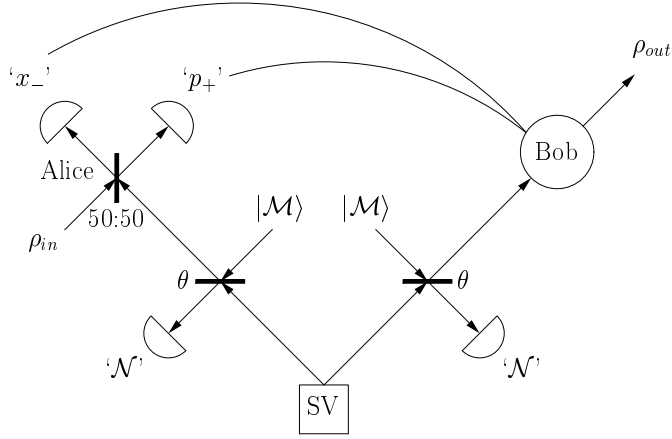, width=\figwidth}}
\caption{Schematic of continuous variable teleportation.  SV is the 
two mode squeezed vacuum entanglement resource, one beam of which goes 
to Alice, the other to Bob.  Alice mixes the unknown input state 
$\rho_{in}$ on the 50:50 beam splitter and measures position 
difference $x_-$, and momentum sum $p_+$.  She sends this information 
to Bob via a classical channel who then makes the relevant unitary 
operations on his beam dependent upon the information from Alice to 
recreate the input state at his location $\rho_{out}$.  The 
conditional resource is made by inserting beam splitters of 
reflectivity $\theta$ in each arm of the teleporter, then putting a 
Fock state $\ket{\mathcal{M}}{}$ at the spare port of the beam 
splitters and detecting $\mathcal{N}$ at the detectors.}
\label{fig:exptlSetup}
\end{figure}

Following Opatrn\'y {\it et al.}, we introduce highly transmitting 
beam splitters into each EPR beam and look for coincidences occurring 
from only one photon being ``shaved'' off each beam.  Such 
coincidences tell us when we have a ``good'' resource and therefore 
when we should teleport, it merely being a matter of time to wait for 
such an occurrence.  We introduce two entanglement resources produced 
by these conditional measurements: the photon subtracted EPR state and 
the photon added EPR state.  Consider the experimental schematic shown 
in \fig{fig:exptlSetup}.
To obtain an expression for the photon subtracted EPR state we calculate 
the effect of introducing a beam splitter into each beam of the EPR 
state 
and expand to second order in the beam splitter reflectivity $\theta$ 
(since $\theta$ is small) and condition on the result $\mathcal{N}=1$ 
at each detector with the vacuum at the spare port of each beam 
splitter ($\mathcal{M}=0$).  The photon subtracted state in the Fock 
basis is
\begin{equation}
\ket{\psi}{p.s.} = \sqrt{\frac{(1-\lambda^2)^3}{1+\lambda^2}} 
\sum_{n=0}^\infty (n+1) \lambda^n \ket{n,n}{AB}
\end{equation}
where $\lambda$ is the squeezing parameter, $A$ and $B$ refer to 
the sender (Alice) and reciever's (Bob's) modes respectively, $p.s.$ 
denotes that this is the photon subtracted resource and we have made 
the definition $\ket{n,n}{AB} \equiv \ket{n}{A}\otimes\ket{n}{B}$.  
The probability of obtaining this state is dependent upon the 
squeezing parameter and the reflectivity of the beam splitter;
\begin{equation}
P(\theta,\lambda) = \theta^4 \frac{1+\lambda^2}{(1-\lambda^2)^3}.
\end{equation}
To find the photon added EPR state we perform the same 
calculation as for the photon subtracted EPR state except 
condition off the result $\mathcal{N}=0$ and have the state 
$\ket{\mathcal{M}=1}{}$ at the 
spare port of each beam splitter.  The photon added state in the Fock 
basis is
\begin{equation}
\ket{\psi}{p.a.} = \sqrt{\frac{(1-\lambda^2)^3}{1+\lambda^2}} 
\sum_{n=0}^\infty (n+1) \lambda^n \ket{n+1,n+1}{AB}
\end{equation}
where $p.a.$ denotes that this is the photon added resource.  This 
state has the same probability of occurring as the photon subtracted 
EPR state.  For brevity, we will concentrate on the photon subtracted 
resource, unless stated otherwise, since the conclusions are similar.

The main drawback of this conditioning technique is the small 
probability of the coincidences occurring.  This, however, is offset 
by the current experimental feasibility of detecting single photon 
coincidences, the knowledge of when to teleport the input as given by 
coincidence events, and the realisation that given finite resources 
(such as squeezing) teleportation can be improved.

\subsection{Conditioning as entanglement distillation}

\begin{figure}
\centerline{\epsfig{file=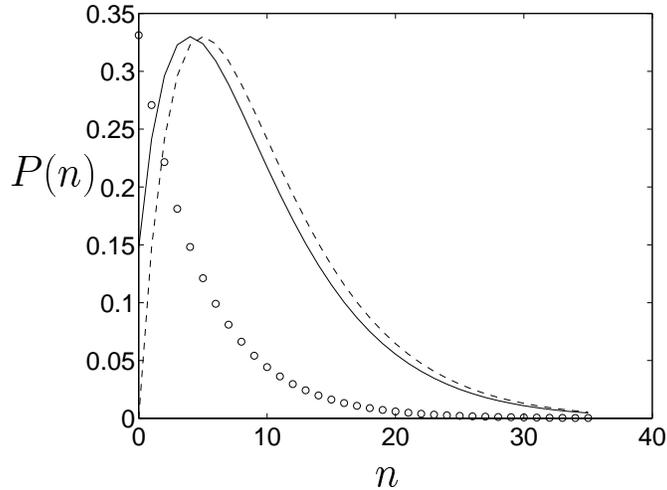, width=\figwidth}}
\caption{Photon number distributions for the standard EPR state 
(circles), the photon subtracted EPR state (solid line) and photon 
added EPR state (dashed line).}
\label{fig:photonDistn}
\end{figure}

\begin{figure}
\centerline{\epsfig{file=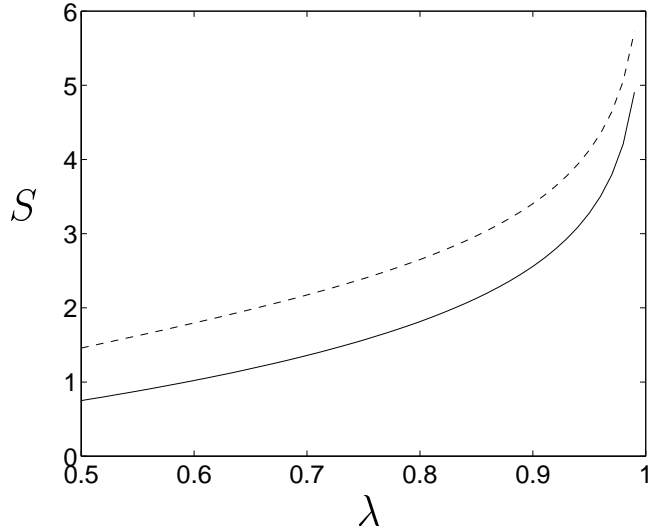, width=\figwidth}}
\caption{Von Neumann entropy, $S$, versus squeezing parameter 
$\lambda$.  The standard EPR state gives the solid curve and the 
photon subtracted EPR state the dashed curve.  The figures shows a 
higher entanglement content in the photon subtracted EPR state 
relative to the standard EPR state.}
\label{fig:entropyLambda}
\end{figure}

The photon number distribution for both the photon subtracted and 
photon added conditional EPR state has a higher weighting for large 
photon numbers than the standard EPR state (see 
\fig{fig:photonDistn}).
This suggests that the conditioning procedure behaves 
similarly to entanglement distillation.  To support this intuition we 
use the fact that the resource states are pure and calculate the von 
Neumann entropy $S = -\mathrm{Tr}(\rho \log \rho)$ as a function of 
the squeezing parameter $\lambda$.  It is well known that the von 
Neumann entropy is a good measure of entanglement for bipartite pure 
states~\cite{Plenio:1998:1}, hence we can analyse the difference in 
entanglement between the standard EPR state and the photon subtracted 
EPR state.  We show in \fig{fig:entropyLambda} the von Neumann entropy 
as a function of squeezing parameter and see that it is higher, given 
a certain level of squeezing, for the photon subtracted EPR state 
(dashed line) than the standard EPR state (solid line).  This result 
shows that the entanglement in the conditional resource is higher than 
that in the standard resource, hence the conditioning procedure seems 
to have the effect of distilling entanglement out of the initial EPR 
state.

\section{Teleportation}

Opatrn\'y {\it et al.}~\cite{Opatrny:2000:1} arrive at expressions for 
the teleportation fidelity, measurement probability and average 
fidelity (the fidelity average over all measurements, weighted by the 
measurement probability) by calculating the effect of the 
teleportation operations on the relevant wavefunctions and then 
transforming into the Fock basis.  In this paper we use the formalism 
of Hofmann~\cite{Hofmann:2000:5} to calculate these parameters.  The 
fidelity is defined as the overlap between the input state 
$\ket{\psi}{T}$ and the output state $\rho_{out}$;
\begin{equation}
F = \bra{\psi}{T} \rho_{out} \ket{\psi}{T}.
\end{equation}
Teleportation in this formalism proceeds as per normal for continuous 
variables~\cite{Braunstein:1998:2,Ralph:1998:1,Furusawa:1998:1,%
Bouwmeester:1997:1}; Alice has one component of an entangled pair of 
states and Bob the other.  She mixes her entangled state with the 
state she wishes to teleport to Bob on a beam splitter, and measures 
the position difference ($x_-$) and momentum sum ($p_+$).  Alice sends 
the measurement result $\beta = x_- + i p_+$ to Bob via a classical 
channel, who now displaces his state by this amount to recreate the 
input state at his location.  The entire teleportation process can be 
described by a transfer operator $\hat{T}(\beta)$ such that
\begin{equation}
\ket{\psi(\beta)}{out} = \hat{T}(\beta) \ket{\psi}{T}
\end{equation}
is the output state, normalised to the probability of measuring the 
result $\beta$,
\begin{equation}
P(\beta) = \braket{\psi(\beta)}{\psi(\beta)}{out}{out}.
\end{equation}
One is able to describe the probability of measuring a 
given $\beta$, the fidelity of teleportation $F(\beta)$ and the 
average fidelity $\bar{F}$, in terms of the transfer operator as 
follows
\begin{eqnarray}
P(\beta) &=& \bra{\psi}{T}\hat{T}^\dagger(\beta) \hat{T}(\beta) 
\ket{\psi}{T},\\
F(\beta) &=& \frac{1}{P(\beta)} \left| \bra{\psi}{T} \hat{T}(\beta) 
\ket{\psi}{T} \right|^2,\\
\bar{F} &=& \int d^2 \beta P(\beta) F(\beta) = \int d^2 \beta \left| 
\bra{\psi}{T} \hat{T}(\beta) \ket{\psi}{T} \right|^2.
\end{eqnarray}
Following this formalism one merely needs to calculate the transfer 
operator for the given entanglement resource in order to obtain the 
parameters of interest.  Hofmann {\it et al.}~\cite{Hofmann:2001:1} 
showed for the standard EPR state that the transfer operator is
\begin{equation}
\hat{T}(\beta) = \sqrt{\frac{1-\lambda^2}{\pi}} \sum_{n=0}^\infty 
\lambda^n \hat{D}_T(g\beta) \ket{n}{}\bra{n}{} \hat{D}_T(-\beta).
\end{equation}
Here $\hat{D}_T(\beta)$ is the displacement of the amount $\beta$, $g$ 
is the gain of the teleporter (to be discussed in more depth in 
\Sec{sec:gain}) and $\lambda$ is the squeezing parameter of the EPR 
state:
\begin{equation}
\ket{\psi}{AB} = \sqrt{1-\lambda^2} \sum_{n=0}^\infty \lambda^n 
\ket{n,n}{AB}.
\end{equation}
By noting correspondences between the standard EPR state and the 
photon subtracted and photon added EPR states one can write 
expressions for the transfer operator for each.  The photon subtracted 
EPR state transfer operator is
\begin{equation}
\hat{T}(\beta) = \sqrt{\frac{(1-\lambda^2)^3}{\pi (1+\lambda^2)}} 
\sum_{n=0}^\infty (n+1) \lambda^n \hat{D}_T(g\beta) \ket{n}{} \bra{n}{} 
\hat{D}_T(-\beta)
\end{equation}
and the photon added EPR state transfer operator is,
\begin{equation}
\hat{T}(\beta) = \sqrt{\frac{(1-\lambda^2)^3}{\pi (1+\lambda^2)}} 
\sum_{n=0}^\infty (n+1) \lambda^n \hat{D}_T(g\beta) \ket{n+1}{} \bra{n+1}{} 
\hat{D}_T(-\beta).
\end{equation}

\section{Position difference, momentum sum teleportation}

\subsection{Teleporting a coherent state}

We consider teleportation of a coherent state to gauge the ability of 
the conditional EPR state relative to the standard EPR state.  The 
fidelity is calculated including a variable gain and output coherent 
amplitude $\gamma$, we do this since we are teleporting coherent 
states and wish to be sufficiently general so as to include the 
possibility of the output state being an attenuated or amplified 
version of the input state.  Using the standard EPR state we find the 
teleportation fidelity for a given measurement $\beta$ to be
\begin{equation}
F(\beta) = e^{-|\gamma - g \beta|^2 - \lambda^2 |\alpha - \beta|^2} 
\left| e^{\lambda (\gamma^* - g \beta^*)(\alpha - \beta)}\right|^2
\end{equation}
with a measurement probability of
\begin{equation}
P(\beta) = \left(\frac{1-\lambda^2}{\pi}\right) e^{-(1-\lambda^2) 
|\alpha - \beta|^2},
\end{equation}
and where the $^*$ superscript denotes the complex conjugate.
For the photon subtracted EPR state the fidelity is
\begin{equation}
F(\beta) = \frac{e^{-|\gamma - g \beta|^2 - \lambda^2 |\alpha - 
\beta|^2}}{\lambda^4 |\alpha - \beta|^4 + 3 \lambda^2 |\alpha - 
\beta|^2 + 1} \left| e^{q(\alpha - \beta)(\gamma^* - g \beta^*)} 
\left[(\alpha - \beta)(\gamma^* - g \beta^*) \lambda + 1\right] \right|^2
\end{equation}
and the measurement probability is
\begin{equation}
P(\beta) = \frac{(1-\lambda^2)^3}{\pi(1+\lambda^2)} e^{(\lambda^2 - 
1) |\alpha - \beta|^2} \left(\lambda^4 |\alpha - \beta|^4 + 3 
\lambda^2 |\alpha - \beta|^2 + 1\right).
\end{equation}
Choosing a coherent state of amplitude $\alpha = 1.5$, an output 
amplitude $\gamma$ also equal to 1.5, unity gain and a squeezing 
parameter value of $\lambda = 0.8$, we find that the average fidelity 
using the standard EPR state is $\bar{F} = 0.9000$, and using the 
photon subtracted EPR state it is $\bar{F} = 0.9246$.  Hence by 
conditioning we can improve the efficacy of teleportation.

\subsection{Teleporting ``cat'' states}

We next turn to the example of teleporting superpositions of coherent 
states of equal amplitude but opposite phase, commonly known as 
``Schr\"odinger cat states''.  Such states are written as
\begin{equation}
\ket{\psi}{cat} = \mathcal{N}_\alpha (\ket{\alpha}{} \pm 
\ket{-\alpha}{}),
\end{equation}
where $\mathcal{N}_\alpha$ is the normalisation given by
\begin{equation}
\mathcal{N}_\alpha = \frac{1}{\sqrt{2 \pm 2 e^{-2|\alpha|^2}}}.
\label{eq:catNorm}
\end{equation}
Even superpositions (the `$+$' form) lead to states of only even 
photon number and are hence referred to as ``even cats'', while odd 
superpositions (the `$-$' form) are referred to as ``odd cats''.  In 
the discussion that follows we will leave the $\pm$ in the equations 
for generality.

Using the Hofmann formalism, the standard EPR state, a gain $g$, an 
input state of amplitude $\alpha$ and a comparison state of amplitude 
$\gamma$ we find the teleportation fidelity to be
\begin{eqnarray}
F(\beta) &=& \mathcal{N}_\alpha^2 \mathcal{N}_\gamma^2 \left( 
\frac{1-\lambda^2}{\pi P(\beta)} \right) \\
&\times &\left| 
e^{-|\gamma - g\beta|^2/2 - |\alpha - \beta|^2/2 + 
\lambda (\gamma^* - g\beta^*) (\alpha - \beta) + 
i \mathrm{Im}(-\gamma g \beta^*) + i\mathrm{Im}(-\beta \alpha^*)} 
\right. \\
&&\pm e^{-|\gamma - g\beta|^2/2 - |\alpha + \beta|^2/2 + 
\lambda (\gamma^* - g\beta^*) (\alpha + \beta) + 
i\mathrm{Im}(-\gamma g \beta^*) + i \mathrm{Im}(\beta \alpha^*)}\\
&&\pm e^{-|\gamma + g\beta|^2/2 - |\alpha - \beta|^2/2 + 
\lambda (\gamma^* + g\beta^*) (\alpha - \beta) + 
i\mathrm{Im}(\gamma g \beta^*) + i \mathrm{Im}(-\beta \alpha^*)}\\
&&+ \left. e^{-|\gamma + g\beta|^2/2 - |\alpha + \beta|^2/2 + 
\lambda (\gamma^* + g\beta^*) (\alpha + \beta) + 
i\mathrm{Im}(\gamma g \beta^*) + i \mathrm{Im}(\beta \alpha^*)}
\right|^2
\end{eqnarray}
with a measurement probability of
\begin{eqnarray}
P(\beta) &=& \mathcal{N}_\alpha^2 \left(\frac{1 - 
\lambda^2}{\pi}\right) \\
&\times&\left[ e^{(\lambda^2 - 1) |\alpha - \beta|^2} \right. \\ 
&&\pm e^{-|\alpha-\beta|^2/2 - |\alpha+\beta|^2/2 - \lambda^2 
(\alpha^*-\beta^*) (\alpha+\beta) + i \mathrm{Im}(-\alpha\beta^*) + 
i \mathrm{Im}(\beta\alpha^*)}\\
&&\pm e^{-|\alpha-\beta|^2/2 - |\alpha+\beta|^2/2 - \lambda^2 
(\alpha^*+\beta^*) (\alpha-\beta) + i \mathrm{Im}(\alpha\beta^*) + 
i \mathrm{Im}(-\beta\alpha^*)}\\
&&+ \left. e^{(\lambda^2 - 1)|\alpha + \beta|^2}\right],
\end{eqnarray}
where $\mathcal{N}_\gamma$ is the normalisation of the comparison 
state, having the same form as \eqn{eq:catNorm}.  Changing the 
entanglement resource to the photon subtracted EPR state we obtain the 
fidelity of teleportation given a result $\beta$,
\begin{eqnarray}
F(\beta) &=& \mathcal{N}_\alpha^2 \mathcal{N}_\gamma^2 \left( 
\frac{(1-\lambda^2)^3}{(1+\lambda^2) \pi P(\beta)} \right) \\
&\times&\left| 
e^{-|\gamma - g\beta|^2/2 - |\alpha - \beta|^2/2 + 
\lambda (\gamma^* - g\beta^*) (\alpha - \beta) + 
i \mathrm{Im}(-\gamma g \beta^*) + i\mathrm{Im}(-\beta \alpha^*)} 
\right. \\
&&\times \left[ (\gamma^* - g\beta^*) (\alpha - \beta) \lambda + 1 
\right]\\
&\pm& e^{-|\gamma - g\beta|^2/2 - |\alpha + \beta|^2/2 + 
\lambda (\gamma^* - g\beta^*) (\alpha + \beta) + 
i\mathrm{Im}(-\gamma g \beta^*) + i \mathrm{Im}(\beta \alpha^*)}\\
&&\times \left[ 1 - (\gamma^* - g\beta^*) (\alpha + \beta) \lambda 
\right]\\
&\pm& e^{-|\gamma + g\beta|^2/2 - |\alpha - \beta|^2/2 + 
\lambda (\gamma^* + g\beta^*) (\alpha - \beta) + 
i\mathrm{Im}(\gamma g \beta^*) + i \mathrm{Im}(-\beta \alpha^*)}\\
&&\times \left[ 1 - (\gamma^* + g\beta^*) (\alpha - \beta) \lambda 
\right]\\
&+& e^{-|\gamma + g\beta|^2/2 - |\alpha + \beta|^2/2 + 
\lambda (\gamma^* + g\beta^*) (\alpha + \beta) + 
i\mathrm{Im}(\gamma g \beta^*) + i \mathrm{Im}(\beta \alpha^*)}\\
&&\times \left. \left[ (\gamma^* + g\beta^*) (\alpha + \beta) \lambda + 1 \right]
\right|^2
\end{eqnarray}
where
\begin{eqnarray}
P(\beta) &=& \mathcal{N}_\alpha^2 
\left(\frac{(1-\lambda^2)^3}{(1+\lambda^2)\pi}\right) \\
&\times&\left[ 
e^{(\lambda^2 - 1) |\alpha - \beta|^2} 
(\lambda^4 |\alpha-\beta|^4 + 3 \lambda^2 |\alpha-\beta|^2 + 1)\right.\\
&\pm& e^{-|\alpha-\beta|^2/2 - |\alpha+\beta|^2/2 - \lambda^2 
(\alpha^*-\beta^*) (\alpha+\beta) + i \mathrm{Im}(-\alpha\beta^*) + 
i \mathrm{Im}(\beta\alpha^*)}\\
&&\times \left(1 + (\alpha+\beta) (\alpha^*-\beta^*) \lambda^2 \left[ 
(\alpha+\beta)(\alpha^*-\beta^*)  \lambda^2 - 3\right]\right) \\
&\pm& e^{-|\alpha-\beta|^2/2 - |\alpha+\beta|^2/2 - \lambda^2 
(\alpha^*+\beta^*) (\alpha-\beta) + i \mathrm{Im}(\alpha\beta^*) + 
i \mathrm{Im}(-\beta\alpha^*)}\\
&&\times \left(1 + (\alpha-\beta) (\alpha^*+\beta^*) \lambda^2 \left[ 
(\alpha-\beta)(\alpha^*+\beta^*) \lambda^2 - 3\right] \right)\\
&+& \left. e^{(\lambda^2 - 1)|\alpha + \beta|^2} 
(\lambda^4 |\alpha+\beta|^4 + 3 \lambda^2 |\alpha+\beta|^2 + 1)
\right].
\end{eqnarray}
For an even cat state of amplitude $\alpha = 1.5$, a squeezing 
parameter $\lambda = 0.8$, a comparison cat state amplitude $\gamma$ 
of 1.5 and unity gain, we find that the standard EPR state has an 
average fidelity of $\bar{F} = 0.6389$ and the photon subtracted EPR 
state has as average fidelity of $\bar{F} = 0.7531$.  Again, the 
conditioning technique improves teleportation efficacy.

Using the result for odd cats, an amplitude of $1.5i$ and a squeezing 
parameter of $\lambda=0.8178$ we can effectively reproduce the results 
of Opatrn\'y {\it et al.}~\cite{Opatrny:2000:1}.  In our case for the 
standard EPR state we obtain an average fidelity of $\bar{F} = 0.6453$ 
and for the conditioned EPR state $\bar{F} = 0.7589$, reproducing the 
result that the conditioned resource does better than the 
unconditioned resource.

\subsection{Varying the gain}
\label{sec:gain}

\begin{figure}
\centerline{\epsfig{file=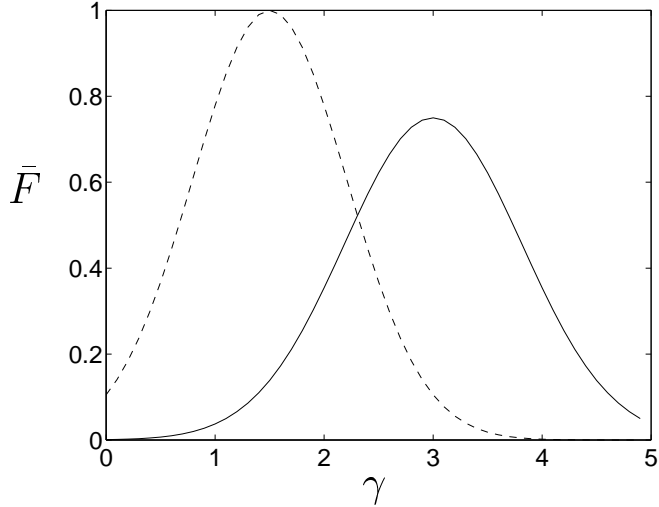, width=\figwidth}}
\caption{Average fidelity as a function of comparison state amplitude 
$\gamma$ for an input coherent state of amplitude $\alpha=3$ using the 
standard EPR state.  The solid curve is the fidelity distribution at 
unity gain and the dashed curve is at gain 0.5.  Note that the 
fidelity goes to 1, but at a reduced amplitude of $\gamma=1.5$.}
\label{fig:gainCohst}
\end{figure}

\begin{figure}
\centerline{\epsfig{file=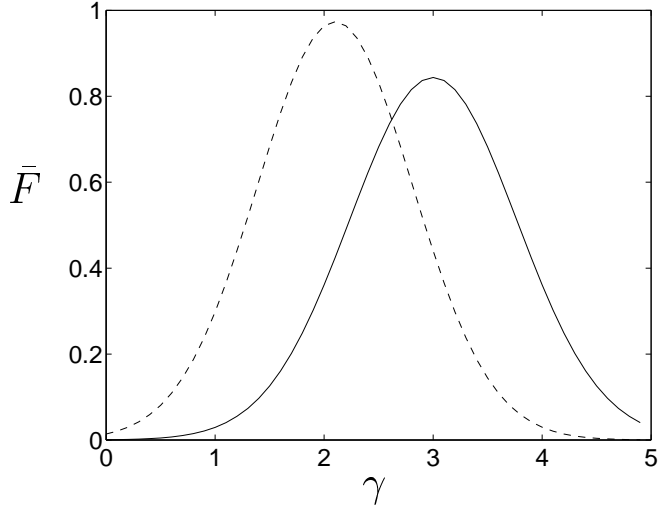, width=\figwidth}}
\caption{Average fidelity as a function of comparison state amplitude 
$\gamma$ for an input coherent state of amplitude $\alpha=3$ using the 
photon subtracted EPR state.  The solid curve is the fidelity 
distribution at unity gain and the dashed curve is at gain 0.7.  The 
fidelity improves by reducing the gain and is a maximum at lower 
amplitude ($\gamma=2.1$) but does quite not reach unity.}
\label{fig:gainCondCohst}
\end{figure}

Polkinghorne and Ralph~\cite{Polkinghorne:1999:1} (and later Hofmann 
{\it et al.}~\cite{Hofmann:2001:1}) identified a particular gain for 
which the output exactly corresponds to an attenuated version of the 
input.  We see the same effect here, for if we use a coherent state of 
amplitude $\alpha=3$ as input, a squeezing parameter of 0.5, and then 
vary the output comparison state, we find that the average fidelity 
indeed goes to 1 (see \fig{fig:gainCohst})
but producing an output state of amplitude $\gamma=1.5$.  This effect 
is also evident if we use the photon subtracted EPR state, 
unfortunately the average fidelity is unable to reach unity, as shown 
in \fig{fig:gainCondCohst}.
For this resource the maximum occurs at a gain of 0.7 and a comparison 
state amplitude of $\gamma = 2.1$.  Note that although the average 
fidelity does not quite reach unity, the attenuated amplitude is 
somewhat higher than that achieved using the standard EPR state, 
implying that the efficiency of the teleporter may have been improved 
as a result of the conditioning procedure.

\subsection{Generalisation}

We wish to briefly note that if one has an entanglement resource of 
the form
\begin{equation}
\ket{\psi}{} = \mathcal{N} \sum_{n=0}^\infty c_n \ket{n,n}{},
\end{equation}
where $\mathcal{N}$ is the normalisation of the entangled state and 
the $c_n$ are the coefficients that describe the photon number 
distribution of the state, one can generalise the Hofmann transfer 
operator to
\begin{equation}
\hat{T}(\beta) = \sqrt{\frac{\mathcal{N}}{\pi}} \sum_{n=0}^\infty 
c_n \hat{D}_T(g \beta) \ket{n}{} \bra{n}{} \hat{D}_T(-\beta).
\end{equation}
It is easy to see that the three entanglement resources discussed in 
this paper are in this form.  The significance of this result is that 
one has some freedom to choose an entanglement resource applying 
directly to the given situation, which may help to enhance the 
teleportation fidelity or ease of implementation of the protocol.

\section{Number difference, phase sum teleportation}

Milburn and Braunstein~\cite{Milburn:1999:1} introduced a 
teleportation protocol using number difference and phase sum 
measurements on the standard EPR state.  Their protocol has the same 
structure as the more usual teleportation scheme involving the two 
mode squeezed vacuum but the measurements made by Alice are of number 
difference and phase sum.  We now show that by making photon 
subtracted and added conditional measurements on the entanglement 
resource improves this protocol also.

\begin{figure}
\centerline{\epsfig{file=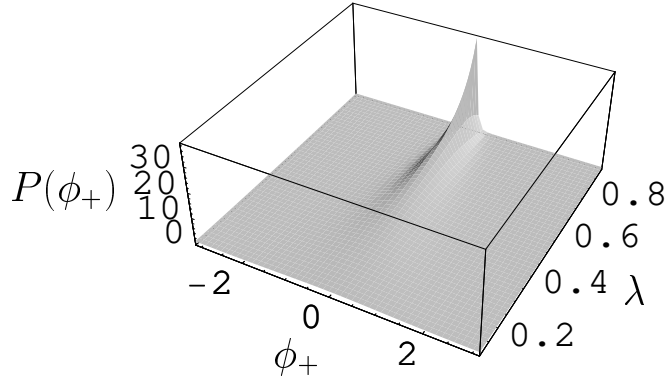, width=\figwidth}}
\caption{Joint phase probability distribution as a function of both 
phase sum $\phi_+$ and squeezing parameter $\lambda$.  The 
distribution becomes sharply peaked with increasing $\lambda$ 
indicating that the photon subtracted EPR state is tending towards 
eigenstates of phase sum.}
\label{fig:jointPhaseDistn}
\end{figure}

The usual EPR state is an eigenstate of number difference and a near 
eigenstate of phase sum for $\lambda$ close to 
unity~\cite{Milburn:1999:1}.  To see that the photon subtracted and 
added EPR states also fulfil these criteria we reiterate their form in 
the Fock basis.  Firstly, the photon subtracted EPR state,
\begin{equation}
\ket{\psi}{p.s.} = \sqrt{\frac{(1-\lambda^2)^3}{1+\lambda^2}} 
\sum_{n=0}^\infty (n+1) \lambda^n \ket{n,n}{}
\end{equation}
and secondly, the photon added EPR state,
\begin{equation}
\ket{\psi}{p.a.} = \sqrt{\frac{(1-\lambda^2)^3}{1+\lambda^2}} 
\sum_{n=0}^\infty (n+1) \lambda^n \ket{n+1,n+1}{}
\end{equation}
These states are obviously eigenstates of number difference, however, 
it is not so easy to see that they are also near eigenstates of phase 
sum.  To see this we calculate the joint phase probability density
\begin{equation}
P(\phi_1,\phi_2) = \left| 
\bra{\phi_1}{}\braket{\phi_2}{\psi}{}{AB}\right|^2
\end{equation}
where the $\ket{\phi_j}{}$ are the phase states
\begin{equation}
\ket{\phi_j}{} = \sum_{n=0}^\infty e^{-i \phi_j n} \ket{n}{}.
\end{equation}
The joint phase probability density for the photon subtracted EPR 
state may be written more explicitly as
\begin{equation}
P(\phi_+) = \frac{(1-\lambda^2)^3}{1+\lambda^2} \left| 
\sum_{n=0}^\infty e^{i n \phi_+} (n+1) \lambda^n \right|^2
\end{equation}
where $\phi_+ = \phi_1 + \phi_2$.  For the photon added EPR state we 
have
\begin{equation}
P(\phi_+) = \frac{(1-\lambda^2)^3}{1+\lambda^2} \left| 
\sum_{n=0}^\infty e^{i (n+1) \phi_+} (n+1) \lambda^n \right|^2.
\end{equation}
As $\lambda \rightarrow 1$ these distributions become more peaked 
about $\phi_+ = 0$ on the range $[\pi,-\pi]$, showing us that the 
phase is highly correlated and the states are close to eigenstates of 
phase sum.  This is shown in \fig{fig:jointPhaseDistn} for the photon 
subtracted case only, since the photon added distribution provides the 
same conclusion.

Teleportation proceeds as described in Refs~\cite{Milburn:1999:1} and 
\cite{Cochrane:2000:2}.  We find for a given (positive) number 
difference measurement, $k$, between Alice's mode and the input state, 
that for the photon subtracted EPR state the state in Bob's mode is
\begin{equation}
\ket{\psi}{out} = \sqrt{\frac{(1-\lambda^2)^3}{(1+\lambda^2) P(k)}}
\sum_{n=0}^\infty c_{n+k} (n+1) \lambda^n \ket{n+k}{B},
\end{equation}
where the $c_n$ are the coefficients describing the photon number 
distribution of the input state and $P(k)$ is the probability of 
measuring the number difference $k$, which is given by
\begin{equation}
P(k) = \frac{(1-\lambda^2)^3}{1+\lambda^2} \sum_{n=0}^\infty 
|c_{n+k}|^2 (n+1)^2 \lambda^{2n}. 
\end{equation}
The teleportation fidelity given $k$ is
\begin{equation}
F(k) = \frac{(1-\lambda^2)^3}{(1+\lambda^2) P(k)} \left| \sum_{n=0}^\infty 
|c_{n+k}|^2 (n+1) \lambda^n \right|^2.
\end{equation}
For the photon added EPR state the state in Bob's mode is
\begin{equation}
\ket{\psi}{out} = \sqrt{\frac{(1-\lambda^2)^3}{(1+\lambda^2) P(k)}} 
\sum_{n=0}^\infty c_{n+k+1} (n+1) \lambda^n \ket{n+k+1}{B}
\end{equation}
where $P(k)$ is 
\begin{equation}
P(k) = \frac{(1-\lambda^2)^3}{1+\lambda^2} \sum_{n=0}^\infty 
|c_{n+k+1}|^2 (n+1)^2 \lambda^{2n} 
\end{equation}
and the teleportation fidelity is
\begin{equation}
F(k) = \frac{(1-\lambda^2)^3}{(1+\lambda^2) P(k)} \left| \sum_{n=0}^\infty 
|c_{n+k+1}|^2 (n+1) \lambda^n \right|^2.
\end{equation}
Reiterating the results for the standard EPR 
state~\cite{Cochrane:2000:2} we have the state in Bob's mode,
\begin{equation}
\ket{\psi}{out} = \sqrt{\frac{1-\lambda^2}{P(k)}} \sum_{n=0}^\infty 
c_{n+k} \lambda^n \ket{n+k}{B}
\end{equation}
with corresponding measurement probability
\begin{equation}
P(k) = (1-\lambda^2) \sum_{n=0}^\infty |c_{n+k}|^2 \lambda^{2n}
\end{equation}
and teleportation fidelity
\begin{equation}
F(k) = \frac{1-\lambda^2}{P(k)} \left| \sum_{n=0}^\infty 
|c_{n+k}|^2 \lambda^n \right|^2.
\end{equation}

To gauge the performance of the entanglement resources we 
consider teleportation of a coherent state of amplitude $\alpha=3$.
Such a choice means the $c_n$ coefficients are given by
\begin{equation}
c_n = e^{-|\alpha|^2/2} \frac{\alpha^n}{n!}
\end{equation}
We only wish to illustrate the relative performance of the resources 
as a function of squeezing parameter, so to simplify the graphical 
output we consider the cases of a number difference measurement of 
$k=0$ and $k=5$.
\begin{figure}
\centerline{\epsfig{file=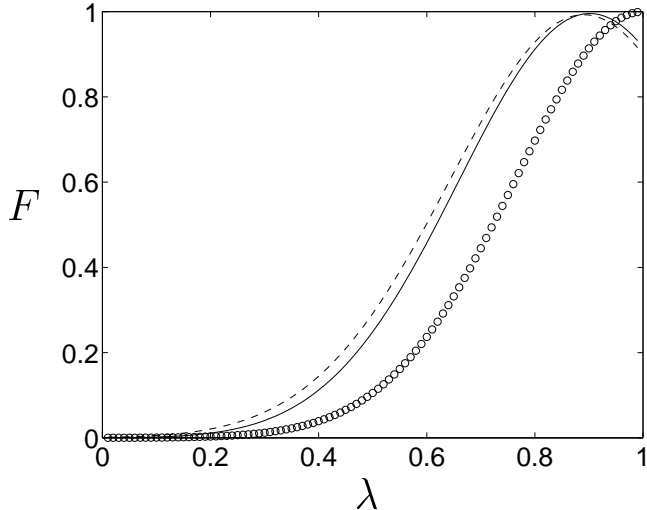, width=\figwidth}}
\caption{Fidelity $F$ as a function of squeezing parameter $\lambda$ 
given a measurement result $k=0$.  The circles denote the distribution 
for the standard EPR state, the solid line for the photon subtracted 
EPR state and the dashed line the photon added EPR state.}
\label{fig:ndiffpsumk0}
\end{figure}
\begin{figure}
\centerline{\epsfig{file=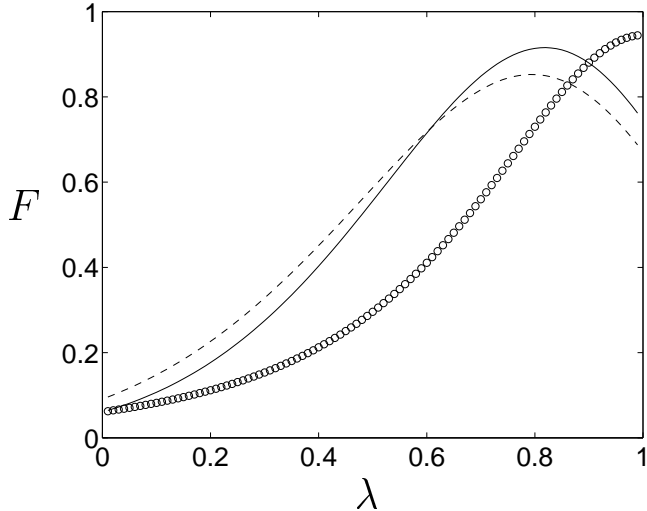, width=\figwidth}}
\caption{Fidelity $F$ as a function of squeezing parameter $\lambda$ 
given a measurement result $k=5$.  The circles denote the distribution 
for the standard EPR state, the solid line for the photon subtracted 
EPR state and the dashed line the photon added EPR state.}
\label{fig:ndiffpsumk5}
\end{figure}
From \fig{fig:ndiffpsumk0} we see that both the photon subtracted 
(solid line) and photon added (dashed line) EPR states have a higher 
fidelity than the standard EPR resource for a large range of the 
squeezing parameter, implying that the conditional resources do a 
better job of teleportation.  The photon added EPR state does slightly 
better than its counterpart for this value of the number difference 
measurement, however as we can see from \fig{fig:ndiffpsumk5}, for 
larger $k$ it does not do as well in comparison to the photon 
subtracted resource, and in fact both conditional EPR states fail to 
achieve as high a fidelity as the more standard resource.  
Nevertheless, using the conditional resource still improves the 
teleportation for a broad range of experimentally realistic levels of 
squeezing.

\section{Beating $\bar{F} = 2/3$ via conditioning}

\begin{figure}
\centerline{\epsfig{file=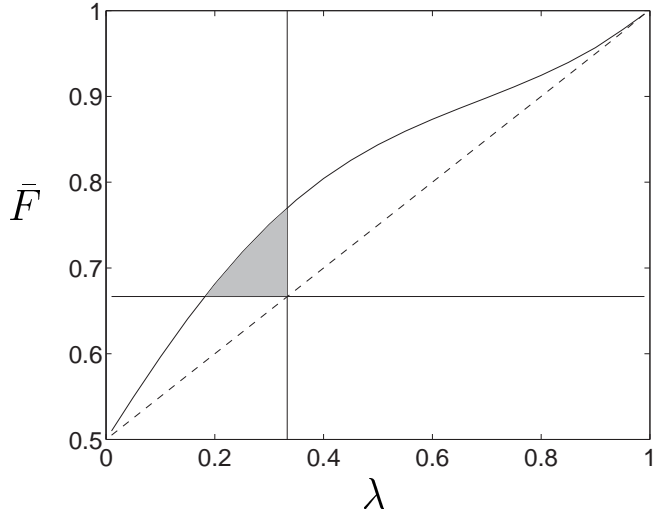, width=\figwidth}}
\caption{Average fidelity $\bar{F}$ as a function of squeezing 
parameter $\lambda$ for teleportation of a coherent state of amplitude 
$\alpha=3$ using the standard EPR state (dashed line) and the photon 
subtracted EPR state (solid line).  The grey shaded region denotes 
where the photon subtracted EPR state beats the 2/3 successful quantum 
teleportation limit whereas the standard EPR state does not.  The 
horizontal line denotes the $\bar{F} = 2/3$ boundary and the vertical 
line gives the right hand edge of the shaded region and is where the 
standard EPR state lies on the boundary.  The maximum difference 
between the two curves occurs at $\lambda=0.37$, giving an improvement 
of 15\%.}
\label{fig:definitelyQuantum}
\end{figure}

The boundary beyond which entanglement is required in continuous 
variable teleportation of coherent states was found by 
Furusawa~\cite{Furusawa:1998:1} to be $\bar{F} = 0.5$.  On the other 
hand a qualitatively different boundary, beyond which the state 
reproduction is unambiguously quantum was found by Ralph and 
Lam~\cite{Ralph:1998:1} and has recently been the source of 
considerable 
discussion~\cite{Ralph:2001:1,Grosshans:2001:1,Braunstein:2001:1}.  
The criterion for beating this second boundary at unity gain was given 
by Ralph and Lam~\cite{Ralph:1998:1,Ralph:2001:1}, and Grosshans and 
Grangier~\cite{Grosshans:2001:1} to be $\bar{F} > 2/3$.  Consider the 
average fidelity of both the standard EPR state and the photon 
subtracted EPR state as functions of the squeezing parameter 
$\lambda$, shown in \fig{fig:definitelyQuantum}, where we teleport a 
coherent state of amplitude $\alpha=3$ with the teleporter at unity 
gain.  We can find a region where the conditional resource beats the 
$2/3$ successful quantum teleportation limit whilst the standard 
resource does not; this region is shaded grey in the figure.
The horizontal line denotes the $\bar{F}=2/3$ boundary and the 
vertical line gives the upper edge of the shaded region and occurs 
where the standard EPR state lies on the $2/3$ boundary.

\section{Discussion}

We have shown how one can improve the efficacy of teleportation by 
making conditional measurements on the Einstein-Podolsky-Rosen 
squeezed vacuum for both the position difference, momentum sum and 
number difference, phase sum continuous variable teleportation 
protocols.  The conditional measurements only require single photon 
coincidence detection which is currently feasible in the laboratory, 
the coincidence events also indicating when is best to teleport.  We 
have also shown that the conditional EPR state gives a resource able 
to provide successfully quantum teleportation for a large range of 
squeezing.

\begin{acknowledgments}
PTC acknowledges the financial support of the Centre for Laser Science 
and the University of Queensland Postgraduate Research Scholarship.  
\end{acknowledgments}


\end{document}